\documentclass[10pt, conference, compsocconf]{IEEEtran}
\usepackage[english]{babel}															% English language/hyphenation
\usepackage[protrusion=true,expansion=true]{microtype}	
\usepackage{amsmath,amsfonts,amsthm} % Math packages
\usepackage[pdftex]{graphicx}	
\usepackage{url}
\usepackage{syntax}
\usepackage{array}
\usepackage{balance}
\usepackage{txfonts}
\usepackage{listings}
\usepackage{verbatim}
\usepackage{todonotes}
\usepackage{xcolor}
\usepackage{url}

\usepackage{fancyhdr}
\newcommand{\jolkw}[1]{\texttt{\bfseries #1}}
\newcommand{\jol}[1]{\texttt{#1}}

\numberwithin{equation}{section}		% Equationnumbering: section.eq#
\numberwithin{figure}{section}			% Figurenumbering: section.fig#
\numberwithin{table}{section}				% Tablenumbering: section.tab#

\definecolor{color:keyword}{rgb}{0.53,0.05,0.05}
\definecolor{color:comment}{rgb}{0.25,0.37,0.75}
\definecolor{color:string}{rgb}{0.87,0.0,0.0}
\usepackage{bold-extra}

\lstdefinelanguage{Jolie}{
morekeywords={
provide,until,OneWay,RequestResponse,new, type, main,define,inputPort,outputPort,init,execution,include,	cset,if,else,csets,interface, throws,global,constants,for,
foreach,while,int,double,raw,void,undefined,string,long,bool,any,single, sequential, concurrent, Jolie, Java, JavaScript, embedded, Location, Protocol, Interfaces, Aggregates, scope, install, cH, comp, throw, this, default, synchronized, nullProcess, false, true},
sensitive=true,
comment=[l]{//},
morecomment=[s]{/*}{*/},
morestring=[b]",
otherkeywords={;,|,:}
}

\lstdefinelanguage{z3}{
morekeywords={
assert, false, true, declare, sort, fun, Bool, String, RegEx, forall, const, define, let, not, implies, iff, and, or},
sensitive=true,
comment=[l]{//},
morecomment=[s]{/*}{*/},
comment=[l]{;},
morestring=[b]",
otherkeywords={|,:}
}

\title{Jolie Community on the Rise}

\author{ 
\IEEEauthorblockN{Alexey Bandura\IEEEauthorrefmark{1},
Nikita Kurilenko\IEEEauthorrefmark{1},
Manuel Mazzara\IEEEauthorrefmark{1},
Victor Rivera\IEEEauthorrefmark{1},
Larisa Safina\IEEEauthorrefmark{1},
Alexander Tchitchigin\IEEEauthorrefmark{1}\IEEEauthorrefmark{2}
}

\IEEEauthorblockA{\IEEEauthorrefmark{1}Innopolis University, Russia
    \\\{a.bandura, n.kurilenko, l.safina, a.chichigin, m.mazzara, v.rivera\}@innopolis.ru}
    
\IEEEauthorblockA{\IEEEauthorrefmark{2}Kazan Federal University, Russia
    \\ a.tchichigin@it.kfu.ru}
}

\begin{document}
\balance 
\maketitle

\begin{abstract}
Jolie is a programming language that follows the microservices
paradigm. As an open source project, it has built a community of
developers worldwide - both in the industry as well as in academia -
taken care of the development, continuously improved its usability,
and therefore broadened the adoption. In this paper, we present some of the most recent results and work in progress that has been made within our research team.
\end{abstract}

\section{Introduction}

Monolithic applications have long been widespread. They are applications 
composed of modules that are not independent  from the application to 
which they belong since they share resources of the same machine (e.g. 
memory, files). Hence, monolithic applications are not dependently executable,
bringing with them a series of drawbacks, e.g. 
scalability and maintainability: a change to a small part of the
application requires the entire monolith to be rebuilt and
redeployed. Plugging together components the same way that
construction engineering does, component-based development appeared as a much more flexible approach. 
To some extent, ``Microservice Architecture'' \cite{fowler} is the extreme of this run towards componentization observed over the last few decades. It differs from a typical Service-Oriented Architecture in terms of a service size, bounded context and independency~\cite{ms-book}. A microservice combines related functionalities into a single business unit implemented as a minimal independent process interacting with other microservices via messages. 
%This makes the architectural design of a system grow up from the structural design of the organization producing it. 
So, instead of having a monolithic
application built as a single unit, the system is built as a set of
small autonomous services that are independently deployable, have firm module boundaries, and can be implemented using a variety of
programming languages and technologies. These services run their own processes and use lightweight communication mechanisms. The term describing the new architectural style has been coined in recent years and has attracted lots of attention as a new way to structure applications.

The Jolie programming language \cite{jolie,jolie:website} was created
in order to maximize the use of the microservices architectural
style. In Jolie, microservices are first-class citizens: every
microservice can be reused, orchestrated, and aggregated with
others~\cite{montesi}. This approach brings simplicity in components
management, reducing development and maintenance costs, and supporting
distributed deployments~\cite{fowler-tradeoffs}.

The development of Jolie followed a major formalization effort for
workflow and service composition languages, and the EU Project
SENSORIA~\cite{sensoria} has successfully produced a plethora of
models for reasoning about composition of services
(e.g.,~\cite{LucchiM07,Mazzara11,mazzaraPhD}). On the mathematical side, the
formal semantics of Jolie~\cite{sock,GLMZ09,MC11} have been inspired
by different process calculi, such as CCS \cite{CCS:Milner80} and the
$\pi$-calculus~\cite{MPW92}. From a practical point of view, however,
Jolie is a descendant of standards for Service-Oriented Computing such
as, for example, WS-BPEL~\cite{bpel}. With both theoretical and
practical influences, Jolie is a suitable candidate for the
application of recent research techniques, e.g., runtime
adaptation~\cite{PGLMG14}, process-aware web applications~\cite{M14},
or correctness-by-construction in concurrent software~\cite{CM13}.

%These qualities made Jolie popular among academic communities around
%the world. 

Our research team published several results to this regard. Recent works were devoted to the extension of Jolie type system~\cite{Safina2016} \cite{Tchitchigin2016}, showing how computation can be moved from a process-driven to a data-driven approach by means of introducing a choice type. This paper presents recent developments for the Jolie language carried out by our research team at Innopolis University. The paper is structured as follows. In Section~\ref{jol}, we briefly review the Jolie programming language in order to simplify the understanding of the forthcoming sections. Section~\ref{foreach} presents an implementation of a new version of the \jolkw{foreach} operator that seeks to simplify the iteration over data structures in Jolie programs. Section~\ref{inline} discusses the implementation of an inline documentation for Jolie developed under the form of an \texttt{Atom} package. Finally, Section~\ref{conclusion} draws some conclusions and looks at the future.

\section{Jolie programming language}
\label{jol}
  
Each Jolie program consists of two parts: behavior and deployment. 

The deployment part contains directives that help the Jolie program to
receive and send messages and be orchestrated among other
microservices. Such directives include:
\begin{itemize}
\item \textit{Interfaces}: sets of operations equipped with
  information about their request (and sometimes response) types
\item\textit{Message types}: can be represented as native types,
  linked types, or undefined
\item \textit{Communication ports}: define how communications with
  other services are actually performed
\end{itemize}

The deployment part is separated from the program behavioral part, so
that the same behavior may be reused later with a different deployment
configuration. 

The behavioral part defines the microservice implementation,
containing both computation and communication expressions. 
Each behavior part has a \jolkw{main} procedure that defines an entry
point of execution. Activities performed by the service can
be composed sequentially, in parallel, or with (input guarded)
non-deterministic choice~\cite{jolie}. 
Execution of behavior part is made by means of standard control flow
statements, like  conditional operators, count-controlled loops and
collection-control loop (\jolkw{foreach} operator) which we will
discuss in more details in Section \ref{foreach}.

Communications in Jolie are type-checked at run-time when a server
receives a message \cite{montesi,nielsen}. Message types are
introduced in the deployment part of Jolie programs by means of the
keyword \jolkw{type} followed by the type identifier and 
definition. Type definition can be native, native with
subtypes, native with undefined subnodes, link type, or can be
undefined (meaning that variable is null until a value is assigned to
it). 

%In addition to commonly used native types like \jolkw{int},
%\jolkw{double} or \jolkw{string}, Jolie also has the following types:
%\begin{itemize}
%\item \jolkw{raw}: used for transmission of raw data streams as byte
 % arrays.
%\item \jolkw{void}: used for indicating that no value is contained by
%  the variable.
%\item \jolkw{any}: any native type initialized with this variable will
%  be accepted. 
%\end{itemize}

Types may have any number of subtypes, also known as subnodes for its
tree-like representation. For
example, we can define the following structure path:
%\begin{lstlisting}[language=C++]
%europe.ireland.city = "Dublin"
%\end{lstlisting}

{\normalsize
  \[
  \begin{array}{l}
    \text{\jol{europe.ireland.city = ``Dublin''}}
  \end{array}
  \]
}
%Structure paths in Jolie are tree-like. 
The structure
%\jol{europe.ireland.city} 
has a root node ``europe'' and a subnode
``ireland'', which has a subnode ``city''. Each node has a
cardinality. If it is not specified, then it defaults to one (meaning
that there is only one entity). %Otherwise, the cardinality is defined
%as the interval with minimum and maximum integer boundaries or by
%means of two shortcuts: ``*'' for any quantity of occurrences and
%``?'' for zero or one occurrence. For example, in the following type
%we define that a country can have any quantity of cities, but only one
%capital:
%\begin{lstlisting}[language=C++]
%type country: string {
%	city*: string
%    capital: string
%}
%\end{lstlisting}
%{\normalsize
%  \[
%  \begin{array}{l}
%    \text{\jolkw{type}\jol{ country:}\jolkw{ string }\jol{\{}}\\
%   \text{\hspace*{1cm}\jol{city*: }\jolkw{string}}\\
%    \text{\hspace*{1cm}\jol{capital:}\jolkw{ string}}
%  \end{array}
%  \]
%}

%\newpage
\section{Extending the semantics of \jolkw{foreach} loops in Jolie}
\label{foreach}

% in Jolie can dramatically reduce unnecessary typing and encapsulate
% the superfluous details promoting usability and clarity in service-
% oriented systems.

%\subsection{The need for foreach loop}
In the Jolie language, to iterate over the values of a variable, it is
necessary to specify the path where the node is. For instance, the
follow code iterates over the values of variable \jol{a.b}: 
{\normalsize
  \[
  \begin{array}{l}
    \jolkw{for} \jol{(i = 0, i $<$ \#a.b, i++)\{}\\
    %\hspace*{1cm}
    \jol{ println\@Console(a.b[i]);}%\\
    \jol{ \}}
  \end{array}
  \]
}
The aid \jol{\#} determines the number of values in a specific
node and \jol{a.b[i]} accesses the ith value of subnode \jol{a.b}. Jolie
does not restrict the user on %the number of values in a
the length of the variable
that he can use. Hence, iterating over longer variables nodes
(e.g. \jol{a.b.c.d.e} or \jol{a.b.c.d.e.f.g}) leads to cumbersome loop
body: firstly, it is needed to type the full variable path every time
an access to a specific values is needed. This is a time-consuming
task, especially when variable paths are too long; secondly, the code
uses a counter to iterate over the array. Hence, it is prone to
error. One needs to continually pay attention to the number of items
in the array while also setting up the index itself. 

%What's more, the behavior layer of a Jolie program doesn't have the notion of a variable scope; therefore, a variable declared in the loop initialization part will still be alive after the loop termination. As such, the counter keeps the value from the last iteration of the loop unless it is explicitly removed or overwritten.

%Keeping in mind these observations, the idiom required to iterate over the values of the node is verbose and unsightly. Furthermore, it is deprived of a good structure, which leads to bugs in those parts of the code. As regards our actual intent, these are certainly low-level concerns.

Jolie offers a workaround to improve the previous code making it less
cumbersome. In Jolie, one variable path can be aliased to a variable,
meaning that a long variable path can be expressed in terms of a short
one.  Aliases are created with the \jol{->} operator, for instance the
code 
{\normalsize
  \[
  \begin{array}{l}
    \jol{var1 -> a.b.c.d[1];}\\
    \jol{var2 -> a.b.c}
  \end{array}
  \]
}
%\begin{lstlisting}[language=C++]
%var -> a.b.c.d[1];
%var -> a.b.c
%\end{lstlisting}
aliased the path variable \jol{a.b.c.d[1]} to \jol{var1} and
\jol{a.b.c} to \jol{var2}. So, the previous example could be rewritten
as

{\normalsize
  \[
  \begin{array}{l}
    \jol{var -> a.b;}\\
    \jolkw{for} \jol{(i = 0, i $<$ \#a.b, i++)\{}\\
    \hspace*{1cm}\jol{println\@Console(var[i]);}\\
    \jol{\}}
  \end{array}
  \]
}
%\begin{lstlisting}[language=C++]
%var -> a.b.c.d[i];
%for(i = 0, i < #a.b.c.d, i++){
%	println@Console(var);
%}
%\end{lstlisting}
Even though the code might be clearer and more readable, aliasing in
Jolie does not bring any performance improvement and the user still
needs to deal with indexes.

%TODO check please my images that I've sent to Victor Rivera. I think the following description is a bit incorrect
Another workaround to tackle this problem is to use the
\jolkw{foreach} operator defined in Jolie. \jolkw{foreach} is defined
to transverse Jolie data structures. The syntax is

{\normalsize
  \[
  \begin{array}{l}
    \jolkw{foreach} \jol{ (nameVar : root)\{}
    %\hspace*{1cm}
    \jol{ //code block}
    \jol{ \}}
  \end{array}
  \]
}
The \jolkw{foreach} operator will iterate over the nodes defined in
\jol{root} executing the internal code block at each iteration. The
main issue with this workaround is that it will iterate over the root
node of a path variable. The code

{\normalsize
  \[
  \begin{array}{l}
    \jolkw{foreach} \jol{ (nameVar : a.b)\{}
    %\hspace*{1cm}
    \jol{ //code block}
    \jol{ \}} 
  \end{array}
  \]
}
will iterate over \jol{a} (as it is the root of \jol{a.b}) instead of
over the subnode \jol{a.b}.

The proposed \jolkw{foreach} seeks to follow a more natural scheme:
``iteration through all elements in a specific (sub)node'' while
hiding to users the heavy burden of dealing with indexes. The syntax
and semantics of the current \jolkw{foreach} operator is extended
following similar semantics as the ones found in other programming
languages, e.g. C++ or Java.

\subsection{Extending the \jolkw{foreach} loop in Jolie}
As stated before, the current Jolie syntax does not provide any
mechanism to transverse a (sub)node in the tree path of a variable. The
proposed extension for the \jolkw{foreach} loop will promote clarity
and usability of the language, enabling users to have a more
structured source code. %The extension takes into account the previous
%observations. 

The extension comprises two current Jolie constructs: \jolkw{foreach}
loops and alias syntax: on one hand, the \jolkw{foreach} loop will
keep an implicit index that will be managed by the language, hiding
the heavy burden to the users; and another hand, it will use alias syntax
eliminating clumsy code brought by common long variable paths, while
encapsulates all low-level details that are irrelevant in our context.

The following shows the syntax extension of the \jolkw{foreach}
construct:
{\normalsize
  \[
  \begin{array}{l}
    \jolkw{foreach} \jol{ (lVariablePath -> rVariablePath)\{}\\
    \hspace*{1cm}\jol{//code block using lVariablePath}\\
    \jol{\}}
  \end{array}
  \]
}
where \jol{lVariablePath} is a fresh variable aliased to the variable
path \jol{rVariablePath}. \jol{rVariablePath} needs to represent a
node in the variable path rather than a specific value. The
\jolkw{foreach} operator will iterate over the values defined in the
node \jol{rVariablePath} executing the code block at each iteration.

The implementation is a syntax sugar for the \jolkw{for} loop presented
previously. It will, internally, translates the previous code to:
{\normalsize
  \[
  \begin{array}{l}
    \jolkw{for} \jol{ (i = 0, i $<$ \#rVariablePath, i++)\{}\\
    \hspace*{1cm}\jol{lVariablePath ->  rVariablePath[i];}\\
    \hspace*{1cm}\jol{//code blockusing lVariablePath}\\
    \jol{\}}
  \end{array}
  \]
}

The \jol{foreach} extension was implemented by implementing a visitor
to transverse Jolie's code AST and generating the previous Jolie
code. The implementation can be found at \cite{github}

Having this new syntax and semantics for \jolkw{foreach},
users can iterate over long path variables without introducing
cumbersome code or having the heavy burden of dealing with indexes.

%There is a restriction on the \jol{rVariablePath}. It cannot be a variable path that ends with an index. For instance, even though the variable path \jol{a.b.c[3]} is a valid path in Jolie, it cannot be used however it points to a value within the node rather than a node itself).

%Introducing a new keyword (e.g foreach) could manifest many name clashes since the keyword could be used in a place of an identifier in legacy code. In our case, Jolie is a relatively young language and still in the process of establishing its own community, so we were free to take the new keyword. The general syntax is the following:

Compared to Java where an enhaced for-loop can't be used to remove or update elements as you traverse a collection, foreach in Jolie are capable to do this:  
{\normalsize
  \[
  \begin{array}{l}
    \jolkw{foreach} \jol{ (var -> node.subnode)\{}\\
    \hspace*{1cm}\jol{var = something}\\
    \jol{\}}
  \end{array}
  \]
}
The last point is that Jolie foreach employs an alias sign ``\jol{->}''instead of a colon symbol used in Java and C++ to separate a target collection and an iteration variable. The colon symbol in Jolie is already used to traverse the subnodes of a particular node (e.g. subnode_1, subnode_2, ... subnode_n) while we are iterating through the values of the specific node (e.g node[0], node[1], ... node[n-1]). Additionally, by employing ``\jol{->}'' we explicitly demostrate that foreach uses an alias syntax while traversing values of a node.

%\newpage
\section{Displaying inline documentation for Jolie}
\label{inline}

%\subsection{Need for inline documentation}
Commonly, developers prefer the use of Integrated Development
Environments (IDEs) since they provide a vast amount of 
functionalities within the same context. They come, for instance, with
the corresponding compilers or interpreters, testing and debugging
tools, code navigation and autocompletion, and so on. 

Jolie's IDE is based on Atom \cite{atom} text editor, which is a
powerful open-source editor. %Atom is a desktop application built with
%web technologies on top of the \texttt{Node.js} \cite{} and it is written in
%\texttt{CoffeeScript} \cite{} and \texttt{Less} \cite{}.
 An advantage
of using an IDE based on Atom, is that one can extend its
functionality by implementing additional packages and coupling them to
the main core. 

Jolie's IDE is still under work in progress. For instance, the IDE does
not support the invoking of a definition of a type or a port
(when the cursor is standing above an identifier) as an inline small window to recall the programmer of it.
This section shows the development of a feature for displaying inline
documentation in Jolie's IDE for Jolie's code. The inline
documentation view was implemented as an Atom's package. The package
shows in a small window brief information (the inline
documentation) about the subject which is selected with the
cursor (see Figure \ref{inline:window}). The idea behind this feature
is for the programmer to have a quick view of the documentation of a
specific identifier without the need of opening the file where the documentation 
is.

\subsection{Deployment part of Jolie}
As stated in Section \ref{jol}, Jolie programs consist of two parts: behavior and
deployment. The support for inline documentation takes into account
the deployment part.

In the deployment part, the communication is supported by communication ports. They define the communication links used by services to exchange data. There
exist two kinds of ports: input and output ports and their syntaxes are similar. Ports are based 
upon the three fundamental concepts: \jolkw{Location}, \jolkw{Protocol} and \jolkw{Interface}. The
following is the syntax for an output port in Jolie:

{\normalsize
  \[
  \begin{array}{l}
    \jolkw{outputPort} \jol{ id \{}
	%\hspace*{1cm}
    \jolkw{Location}\jol{: URI}\\
	\hspace*{3cm}\jolkw{Protocol}\jol{: p}\\
	\hspace*{3cm}\jolkw{Interfaces}\jol{: iface\_1}
    \jol{\}}
  \end{array}
  \]
} 

\jol{id} is the name of the port. \jolkw{Location}  expresses the communication medium, \jol{URI} stands for Uniform Resource Identifier, and it defines the location of the port. The \jolkw{protocol} defines how data needs to be sent or received. It is defined by \jol{p}. Finally, \jolkw{Interfaces}  are the interfaces accessible through the port. The implementation of the inline doc view will give
information about the kind of protocols (defined in \jolkw{protocol}) and the interfaces
(defined in \jolkw{Interfaces})

\subsection{Implementation}
\begin{center}
\begin{figure}[t]
	\centering
  \includegraphics[scale=0.4]{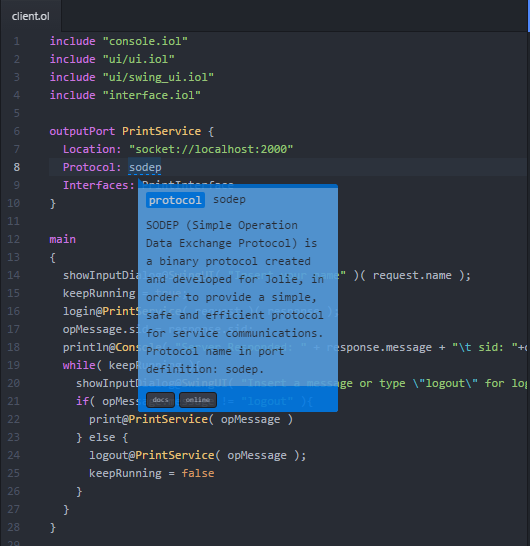}
  \caption{Inline documentation for Jolie code}
  \label{inline:window}
\end{figure}
\end{center}

The inline documentation for Jolie (herein \textit{InlineDoc View}) was implemented as a plugin
of Jolie's IDE (i.e. a package of Atom). When the user clicks on the word he wants
to see the documentation, a window appears showing the corresponding inline
documentation, as depicted in Figure \ref{inline:window}. The window gives a 
snipped of the documentation and presents 
% * <stud.kurylenko@gmail.com> 2016-09-23T14:52:23.467Z:
%
% > snipped
%
% snippet?
%
% ^.
two buttons, ``docs'' and ``online''. ``docs'' gives the user the possibility 
to show the documentation \ldots %TODO
%kurilenko: in an view appearing in the workspace opening exhaustive documentation for the Jolie language regarding the word.
; the ``online'' button translates the representation of the documentation to HTML
code so the documentation can be showed in a browser. 
 
The plugin implements a series of components that interact to each other 
sequentially. When the user clicks on a word in the Jolie IDE, one of
the predefined Atom's components retrieves the  string text
corresponding to the line where the cursor is on. This information is
sent to a 
\texttt{pre-process} component. This component 
\begin{enumerate}
\item filters the line: checks whether the user clicked on a word that
  might contain documentation.  The implementation of the InlineDoc
  View takes into account only the deployment part of Jolie
  programs. If users click on something different, no information is
  retrieved; it also checks whether the word has some meaning, e.g. it
  is not a curly bracket or a Jolie keyword, from which no information
  is available.
\item categorizes the type of word the user clicked on. The word can
  correspond to the protocol or to the interface parts of the port
  communication.
\end{enumerate}

This component sends both the corresponding \textit{word} (which
correspond to a word that some information can be retrieved from) and
the \textit{categorization} (either the word is defined in the
protocol or in the interface part) to the \texttt{evaluation}
component. This
component will search for the documentation of the \textit{word} in the
\textit{categorization} files. Categorization files are JSON files
provided by us that define a mapping between protocols and
documentation. This component will send the corresponding
documentation to another Jolie IDE component that is in charge of
displaying the text: the InlineDoc View window. The information
received by this component uses \texttt{markdown} format
\cite{markdown}. \texttt{Markdown} is a text-to-HTML conversion tool that
enables users to get a structured valid XHTML (or HTML) from a
text. So this component can also display the information (upon request
of the user by clicking on the ``online'' button) in a browser.
%(~\ref{inline:window})
%The process followed for the InlineDoc View is depicted in Figure \ref{process}. The figure shows the
%Jolie's IDE, which is based on Atom. Components (depicted as squares) represent
%the different functionalities. The InlineDoc View was implemented as a 
%package (depicted in Figure \ref{process} as a dotted square).
%\newpage
\section{Conclusive Remarks}
\label{conclusion} 

%Jolie is a programming language that is gaining international attention for its simplicity of use and focus on componentization and orchestration.

Research on microservices and Jolie is progressing and several scientists have noticed the interesting features and the challenges connected with language development. This paper reports our latest results and highlights realization details. Two aspects have been covered here and represent the contribution of our work:

\begin{itemize}
\item \textit{Foreach operator}: we presented an implementation of a new version of the \jolkw{foreach} operator aiming at simplifying the iteration over data structures in Jolie programs. It is sintactic sugar, but makes non-trivial programs more straightforward to write and read.
\item\textit{Inline documentation }: we introduced the implementation of an inline documentation for Jolie developed under the form of an \texttt{Atom} package. Jolie's IDE is work in progress, and much as to be done yet. Still, inline documentation supports developers in writing programs and ultimately enhance code quality.
\end{itemize}

%At the same time that this paper is getting published, 

Open challenges are related to the static type checking of Jolie, a feature that is necessary for a broader adoption of the language in an industrial context. The principal effort of our research in the upcoming months will therefore focus on static analysis.

\small

\section*{Acknowledgements}
The authors would like to thank Daniel Johnston for assistance. Fabrizio Montesi played an important role in providing us with interesting challenges and following the technical execution. In particular, the \jolkw{foreach} extension comes from his specification, and he also contributed to the final realization. The refinement and stabilization of this extension for release is now under way.

\bibliographystyle{plain}
\bibliography{biblio}

\end{document}